\documentclass[prl,aps,twocolumn,showpacs,superscriptaddress,amsmath,amssymb]{revtex4-1}

\usepackage[pdftex]{graphicx}
\usepackage{color}
\definecolor{dred}{rgb}{0.7,0.0,0.0}
\usepackage{dcolumn}
\usepackage{bm}
\allowdisplaybreaks 
\usepackage{braket}
\usepackage{ulem} 
\usepackage{hyperref}
\hypersetup{colorlinks=true, linkcolor=blue, citecolor=blue, filecolor=blue, urlcolor=blue, pdftitle=, pdfauthor=, pdfsubject=, pdfkeywords=}

\newcommand{\bs}[1]{\ensuremath{\boldsymbol{#1}}}

\renewcommand{\i}{\text{i}}

\newcommand{\x}{\text{x}}
\newcommand{\y}{\text{y}}
\newcommand{\z}{\text{z}}

\def\ie{{i.e.},\ }
\def\eg{{e.g.}\ }
\def\etal{{et al.}}

\newcommand {\bu}{{\bar u}}
\newcommand {\bv}{{\bar v}}

\newcommand{\up}{\uparrow}
\newcommand{\dw}{\downarrow}

\newcommand{\parder}[2]{\frac{\partial #1}{\partial #2}}

\definecolor{orange}{rgb}{1,0.5,0}
\definecolor{black}{rgb}{0,0,0}

\begin{document}

\title{Interacting surface states of three-dimensional topological insulators}

\author{Titus Neupert}
\affiliation{Princeton Center for Theoretical Science, Princeton University, Princeton,
New Jersey 08544, USA}

\author{Stephan Rachel}
\affiliation{Institute for Theoretical Physics, Technische Universitaet Dresden, 01171 Dresden, Germany}

\author{Ronny Thomale} \affiliation{ Institute for Theoretical Physics, 
  University of W\"{u}rzburg, Am Hubland, D-97074 W\"{u}rzburg, Germany}

\author{Martin Greiter} \affiliation{ Institute for Theoretical Physics, 
  University of W\"{u}rzburg, Am Hubland, D-97074 W\"{u}rzburg, Germany}

\date{\today}

\begin{abstract}

  We numerically investigate the surface states of a strong
  topological insulator in the presence of strong electron-electron
  interactions.  We choose a spherical topological insulator geometry
  to make the surface amenable to a finite size analysis.  The
  single-particle problem maps to that of Landau orbitals on the
  sphere with a magnetic monopole at the center that has unit strength
  and opposite sign for electrons with opposite spin. Assuming
  density-density contact interactions, we find superconducting and
  anomalous (quantum) Hall phases for attractive and repulsive
  interactions, respectively, as well as chiral fermion and chiral
  Majorana fermion boundary modes between different phases.  Our setup
  is preeminently adapted to the search for topologically ordered
  surface terminations that could be microscopically stabilized by
  tailored surface interaction profiles.

\end{abstract}

\maketitle

{\it Introduction.}---%
Three-dimensional topological
insulators (3DTIs)~\cite{Roy2009,Moore2007,Fu2007,Fu2007-2,Hasan2010,Qi2011}, since their prediction in 2007, realize
a quantum state of matter which triggered enormous interest in condensed
matter physics, and have been
subsequently discovered in various material
classes~\cite{Hsieh2008,Hsieh2009,Xia2009,Zhang2009,Chen2009}.

When viewed as a symmetry-protected topological phase, 3DTIs exhibit a
gapped bulk with two-dimensional gapless edge states protected by U(1)
electron number conservation and time reversal symmetry (TRS),
forbidding any adiabatic deformation into a trivial insulator. The
effective electromagnetic field theory characterizing the 3DTI
contains a topological axion term, with quantized coefficient
$\theta=\pi$ for fermionic 3DTIs as compared to $\theta=0$ for TRS
trivial insulators~\cite{Qi2008}.  On the surface of a 3DTI, where
$\theta$ changes from 0 to $\pi$, the gauge invariance of this
topological field theory demands the presence of a half-integer
Chern--Simons term at the surface.  In the absence of symmetry
breaking, this Chern--Simons term is precisely what allows a single
Dirac cone surface state to be gauge invariant and hence consistently
defined on the 3DTI surface, whereas a single massless Dirac field in
odd space-time dimensions would usually exhibit a parity
anomaly~\cite{Niemi1983,Redlich1984,PhysRevB.88.085104}.

When the protecting U(1) particle number symmetry is broken, such as
by a superconducting proximity effect, the 3DTI surface yields an
unconventional gapped $s$-wave superconductor with Majorana modes in
its vortex cores~\cite{Fu2008}. Upon breaking TRS, such as by a
magnetic coating on the surface, the single surface Dirac cone gaps
out, and the Chern-Simons boundary term of the axion bulk action
manifests itself as a $\nu=1/2$ quantum Hall effect~\cite{Qi2008}
without fractionalized excitations.
The axion term implies the
Witten effect~\cite{Witten1979} by which a odd-half integer charge
binds to magnetic monopoles in the bulk of a 3DTI (see also
e.g. Ref.~\onlinecite{max1}).

All aforementioned properties of 3DTIs do not involve interactions
in the bulk or at the surface. Assuming that the gapped 3DTI bulk is
negligibly renormalized by interactions, it remains to be investigated
how interactions could affect the 3DTI surface. To begin with,
interactions could contribute to breaking the protecting symmetries
explicitly or spontaneously. 
Transcending the mean-field picture, however, interactions could also
give rise to a gapped surface state with intrinsic topological order,
allowing a new kind of phase to enter the realm of competing quantum
states of matter on a 3DTI surface.  Investigations of bosonic 3DTI
surface states established that such gapped surface states in the
absence of symmetry breaking are indeed possible for certain kinds of
topological
order~\cite{Ashvin2013,max1,WangSenthil2013,Burnell2013,motru}. Soon
thereafter, this idea has been formulated for the physically more
relevant fermionic
analogue~\cite{max2,Wang2013,Bonderson2013,Fidkowski2013}.  All these
conceptually important works rely on consistency arguments on the
level of topological field theories and constructions that employ
contrived exactly soluble models. What type of physically attainable
Hamiltonians would exhibit these exotic ground states remains a
challenging question.~\cite{Mross14}

Haldane~\cite{Haldane14} has recently pointed out that, as the
topological surface state only has support in a 2D $k$-space region
with an area $A_k$ that may be much smaller that the Brillouin zone,
the surface electrons obey an ``uncertainty principle'' where they
cannot be localized within an area smaller than $(2\pi)^2/A_k$,
analogous to the ``magnetic area'' $h/|eB|$ for electrons confined to
a 2D Landau level.  
He
noted that this makes
the surface dynamics insensitive to the atomic-scale features of the
surface, and leads to a continuum ``fuzzy quantum geometry''
description in which exact diagonalization (ED) studies of
strongly-interacting systems becomes practical, as in the fractional
quantum Hall effect.  Here we will describe an implementation of this
idea using the spherical geometry, instead of the periodic (torus)
geometry of Ref.~\cite{Haldane14}. As a consequence of the inherent
``fuzziness'', one might expect that, in the presence of interactions,
featureless quantum liquid ground states may be favored over
translational-symmetry-breaking order such as lattice-scale
antiferromagnetism or charge-density waves.

In this Letter, we develop a microscopic framework for numerically
studying strong interaction effects on 3DTI surfaces. While a consensus
regarding possible states of matter consistent with the constraints
given by the 3DTI setup is emerging, little is known which of these
states might actually be realized for which kind of interaction
profile. For this purpose, we investigate the surface states of a 3DTI
on a sphere~\cite{Imura12}, the geometry where these states have the
maximally attainable symmetry but no boundary.  The single Dirac cone
at the surface is intimately related to the Landau level quantization
on the sphere~\cite{Greiter11}, with an individual magnetic monopole
of unit (anti-)charge for spin up (down) electrons at the center of
the sphere, so as to respect TRS.  We study the effect of a
density-density contact interaction $U$. We find an $s$-wave
superconducting phase triggered by attractive $U$ as a consequence of
spontaneous U(1) symmetry breaking. For repulsive $U$, the gapless
Dirac cone persists in an extended regime of parameter space. In the
limit of strong interaction, however, we find ferromagnetic phases of
broken TRS. These are the $\nu=1/2$ anomalous quantum
Hall effect and the gapless anomalous Hall effect for fillings at and
away from the Dirac point, respectively.

{\it 3DTI surface states on the sphere.}---In the limit of long wavelengths, the surface states of a strong  3DTI 
are described by a two-dimensional Dirac equation given by
\begin{align}
 \label{eq:3DTI}
  H
  &=v \hat{\bs{n}}\left(-\i\nabla\times\bs{\sigma}\right)
\end{align}
where $v$ denotes the Dirac velocity of the surface states,
$\hat{\bs{n}}$ is the surface normal, and
$\bs{\sigma}=(\sigma_\x,\sigma_\y,\sigma_\z)$ twice the physical
electron spin vector.  For a spherical TI with radius $R$, Imura
\etal~\cite{Imura12} derived that \eqref{eq:3DTI} becomes
\begin{align}
  \label{eq:H0}
  H_0&=\frac{v}{R}\left(\sigma_\x\Lambda_\theta+\sigma_\y\Lambda_\varphi\right)
\end{align}
where 
\begin{align}
  \label{eq:lambda}
  \bs{\Lambda}\;=\;-i\left[
    \bs{e}_\varphi\parder{}{\theta}
    -\bs{e}_\theta\frac{1}{\sin\theta}
    \left(\parder{}{\varphi}-\frac{\i}{2}\sigma_\z\cos\theta\right)
  \right]
\end{align}
is the dynamical angular momentum of an electron in the presence of a
magnetic monopole with strength $2\pi\sigma_\z$, and
$(r,\theta,\varphi)$ are spherical coordinates.  The monopole strength
or Berry flux through the sphere is hence $2\pi$ for $\up$ spins (\ie
spins pointing in $\bs{e}_\text{r}$ direction) and $-2\pi$ for $\dw$
spins (\ie spins pointing in $-\bs{e}_\text{r}$ direction)
\cite{note}.  The origin of this Berry phase is easily understood.
Since the coordinate system for our spins (to which our Pauli matrices
$\sigma_\x,\sigma_\y,\sigma_\z$ refer to) is given by
$\bs{e}_\varphi,-\bs{e}_\theta,\bs{e}_\text{r}$, it will rotate as the
electron is taken around the sphere.  For general trajectories, the
Berry phase generated by this rotation is given by $\frac{1}{2}$ times
the solid angle subtended by the trajectory.  Formally, this phase is
generated by a monopole with strength $2\pi$ at the origin.  Since the
model preserves time reversal invariance, the monopole must be of
opposite sign for opposite spins.

Substitution of \eqref{eq:lambda} into \eqref{eq:H0} yields
\begin{align}
  \label{eq:h0}
  H_0&=\frac{v}{R}\,h_0,\quad
  h_0=\!\left(\!\!
    \begin{array}{cc}
      0&h^+\\h^-&0
    \end{array}\!\!\right)\!,
\end{align}
with 
\begin{align}
  \label{eq:h0+-}
  h^\pm = {\mp\left(\partial_\theta+\frac{1}{2}\cot\theta\right)
    +\frac{\i\partial_\varphi}{\sin\theta}}.
\end{align}
Eq.~\eqref{eq:h0} describes a Dirac hamiltonian in the sense that 
\begin{align}
  \label{eq:h0^2}
  h_0^2=\!\left(\!\!
    \begin{array}{cc}
      h^+h^-&0 \\0&h^-h^+
    \end{array}\!\!\right)
  =\!\left(\!\!
    \begin{array}{cc}
      \bs{\Lambda}^2_{s_0=+\frac{1}{2}}&0 \\
      0&\bs{\Lambda}^2_{s_0=-\frac{1}{2}}
    \end{array}\!\!\right) +\frac{1}{2}
\end{align}
is diagonal.  Apart from an overall numerical factor,
\begin{align}
  \label{eq:qhslambda^2}
  \bs{\Lambda}^2_{s_0}=
     -\frac{1}{\sin\theta}\partial_\theta
     \left(\sin\theta\,\partial_\theta\right)
     -\frac{1}{\sin^2\theta}
     \left(\partial_\varphi-\i s_0\cos\theta\right)^2
\end{align}
is the hamiltonian of an electron moving on a sphere with a monopole
of strength $4\pi s_0$ in the center~\cite{Greiter11}.  The Landau
levels on the sphere are spanned by two mutually commuting SU(2)
algebras, one for the cyclotron momentum ($\bs{S}$) and one for the
guiding center momentum ($\bs{L}$). The Casimir of both is given by
\begin{equation}
  \label{eq:s^2}
  \bs{L}^2=\bs{S}^2=s(s+1),
\end{equation}
where $s=|s_0|+n$ and $n=0,1,\ldots$ is the Landau level index.  With
$\bs{\Lambda}^2=\bs{L}^2-s_0^2=(n+1)^2-\frac{1}{2}$ for
$|s_0|=\frac{1}{2}$, we see that the eigenvalues of $h_0^2$ are
given by $\epsilon^2=(n+1)^2$.

\begin{figure}[t]
\includegraphics[width=0.5\columnwidth]{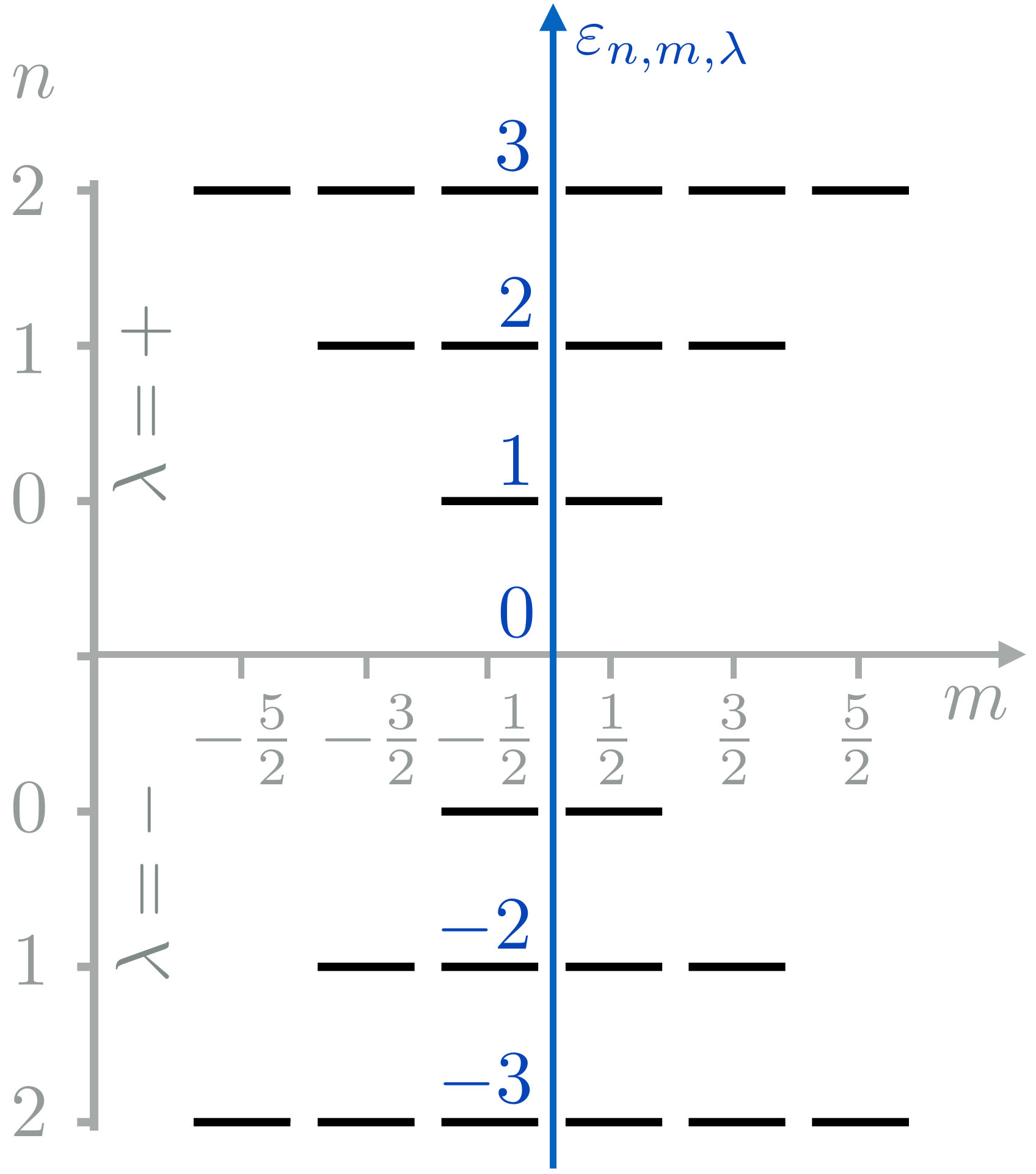}
\caption{(Color online) Single-particle spectrum
  of~\eqref{eq:h0withS+S-} that describes the surface states of a
  spherical topological insulator, including the 24 states closest to
  the Dirac point. The single-particle quantum numbers are the
  ``Landau level" index $n=0,1,2,\cdots$, the angular momentum
  $m=-(n+1/2),\cdots, (n+1/2)$, and the particle-hole index
  $\lambda=\pm$.  }
\label{fig:berry}
\end{figure}

In terms of the spinor coordinates $u=\cos\frac{\theta}{2}
e^{i\frac{\varphi}{2}}$, $v=\sin\frac{\theta}{2}
e^{-i\frac{\varphi}{2}}$ 
introduced by Haldane \cite{Haldane83}, and their complex
conjugates $\bar u$, $\bar v$, 
\begin{align}
  \label{eq:Sdef}
  S^x + iS^y = S^+&=u\partial_\bv-v\partial_\bu,\nonumber\\ 
  S^x - iS^y = S^-&=\bv\partial_u-\bu\partial_v,\\\nonumber
  S^\z &= \textstyle\frac{1}{2}
  (u\partial_u + v\partial_v - \bu\partial_\bu - \bv\partial_\bv),\\[2pt]
  \label{eq:Ldef}
  L^x + iL^y = L^+&=u\partial_v-\bv\partial_\bu,\nonumber\\ 
  L^x - iL^y = L^-&=v\partial_u-\bu\partial_\bv,\\\nonumber
  L^\z &= \textstyle\frac{1}{2}
  (u\partial_u - v\partial_v - \bu\partial_\bu + \bv\partial_\bv).
\end{align}
The physical Hilbert space is restricted to states with $S^\z$
eigenvalue $s_0$, $S^\z\phi=s_0\phi$~\cite{Greiter11}.  With the $\up$
and $\dw$ spin components of the eigenstates of $h_0$ thus restricted
respectively (\ie $S^\z\phi^\up=\frac{1}{2}\phi^\up$ and
$S^\z\phi^\dw=-\frac{1}{2}\phi^\dw$), it is easy to show that
$h^-\phi^\up=-S^-\phi^\up$ and $h^+\phi^\dw=-S^+\phi^\dw$, and hence
that
\begin{align}
  \label{eq:h0withS+S-}
  h_0=\left(\!\!
    \begin{array}{c@{\hspace{2pt}}c} 0&-S^+\\[2pt]-S^-&0 \end{array}
    \!\!\right)\!.
\end{align}
The Dirac property of $h_0$ and the eigenvalues of $h_0^2$ imply that the
eigenstates take the form 
\begin{align}
  \label{eq:psilambda}
  h_0\psi^\lambda_{nm}=\lambda (n+1)\psi^\lambda_{nm},\ \
  \psi^\lambda_{nm}=\left(\!
    \begin{array}{c}\phi^\up_{nm}\\[2pt] \lambda\phi^\dw_{nm}\end{array}
    \!\right)\!,\
\end{align}
where $\lambda=\pm 1$ distinguishes positive and negativ energy
solutions, and $m$ is the eigenvalue of $L^\z$.  With
$h^+h^-=S^-S^++1$, we find~\cite{Greiter11}
\begin{align}
  \label{eq:phiup}
  \phi^\up_{nm}=(L^-)^{s-m}(S^-)^n u^{2s}
  =(L^-)^{s-m}\,{\bar v}^n u^{n+1},
\end{align}
where $s=n+\frac{1}{2}$ and $m=-s,-s+1,\ldots ,s$.  With
\eqref{eq:psilambda},
\begin{align}
  \label{eq:phidw}
  \phi^\dw_{nm}=-\frac{S^-}{n+1}\phi^\up_{nm} 
  =-(L^-)^{s-m}\,u^n{\bar v}^{n+1}.
\end{align}
The number of degenerate states in the $(n+1)$-th Landau level with
energy $\epsilon=\lambda (n+1)$ is hence $2(n+1)$, and grows linearly with
$|\epsilon|$, as required for a Dirac cone (see Fig.~\ref{fig:berry}).

$H_0$ is invariant under both time reversal
$\text{T}\equiv-\i\sigma_\y\text{K}$ (where K denotes complex conjugation)
and parity $\text{P}\equiv\sigma_\x\text{P}_\theta$ (where
$\text{P}_\theta$ takes $\theta\to\pi-\theta$).  The basis states 
\eqref{eq:psilambda} transform according to
\begin{align}
  \label{eq:TPpsilambda}
  \text{T}\psi^\lambda_{nm}&=\lambda\, (-1)^{m-\frac{1}{2}}\,\psi^\lambda_{n,-m},\\
  \text{P}\psi^\lambda_{nm}&=\lambda\, (-1)^{n+m+\frac{1}{2}}\,\psi^\lambda_{n,m}.
\end{align}

{\it Momentum space cutoff.}---The Dirac Hamiltonian $H_0$ (or $h_0$)
governs the behavior of the surface states of a topological insulator
for energies close to the Dirac nodal point.  At higher and lower
energies, the surface states merge with the bulk conduction and
valence bands, respectively, and their weight on the surface
diminishes.  Consequently, even strong electron-electron interactions
of the order of the bulk gap will only induce small matrix elements
between bulk and surface states. It is hence sensible to study the
effects of strong interactions on the surface states alone, when
working in the Fock space constructed from the single-particle
eigenstates of $H_0$ with $n\leq n_0$ for some Landau level cutoff
$n_0$.  Importantly, it is impossible to build orbitals in position
space that are localized on length scales smaller than $2\pi R/n_0$ in
this restricted Hilbert space. Thus even if the interactions are much
larger than the kinetic energy scale $v/R$, the problem does not
reduce to a classical limit~\cite{Haldane14}. This is somewhat
reminiscent of the Landau level problem, with the important difference
that single-particle states are exponentially localizable on long
enough distances on the topological insulator surface while they are
power-law decaying in a Landau level on a compact manifold.

{\it Interactions.}---On this restricted single particle Hilbert
space, we consider a contact interaction
\begin{equation}
  H_{\mathrm{int}}
  =U\int_{S_2}\mathrm{d}^2\;\bs{r}\rho_\downarrow(\bs{r})\rho_\uparrow(\bs{r}),
\label{eq: H int}
\end{equation}
where $\rho_s(\bs{r})$ is the density operator of electrons with spin
$s$ at position $\bs{r}$.  This interaction preserves T, P, the
number of particles $N_{\mathrm{p}}$, and the total angular momentum
$M=\sum_{i=1}^{N_{\mathrm{p}}}m_i$.
\begin{figure}[t]
\includegraphics[width=0.98\columnwidth]{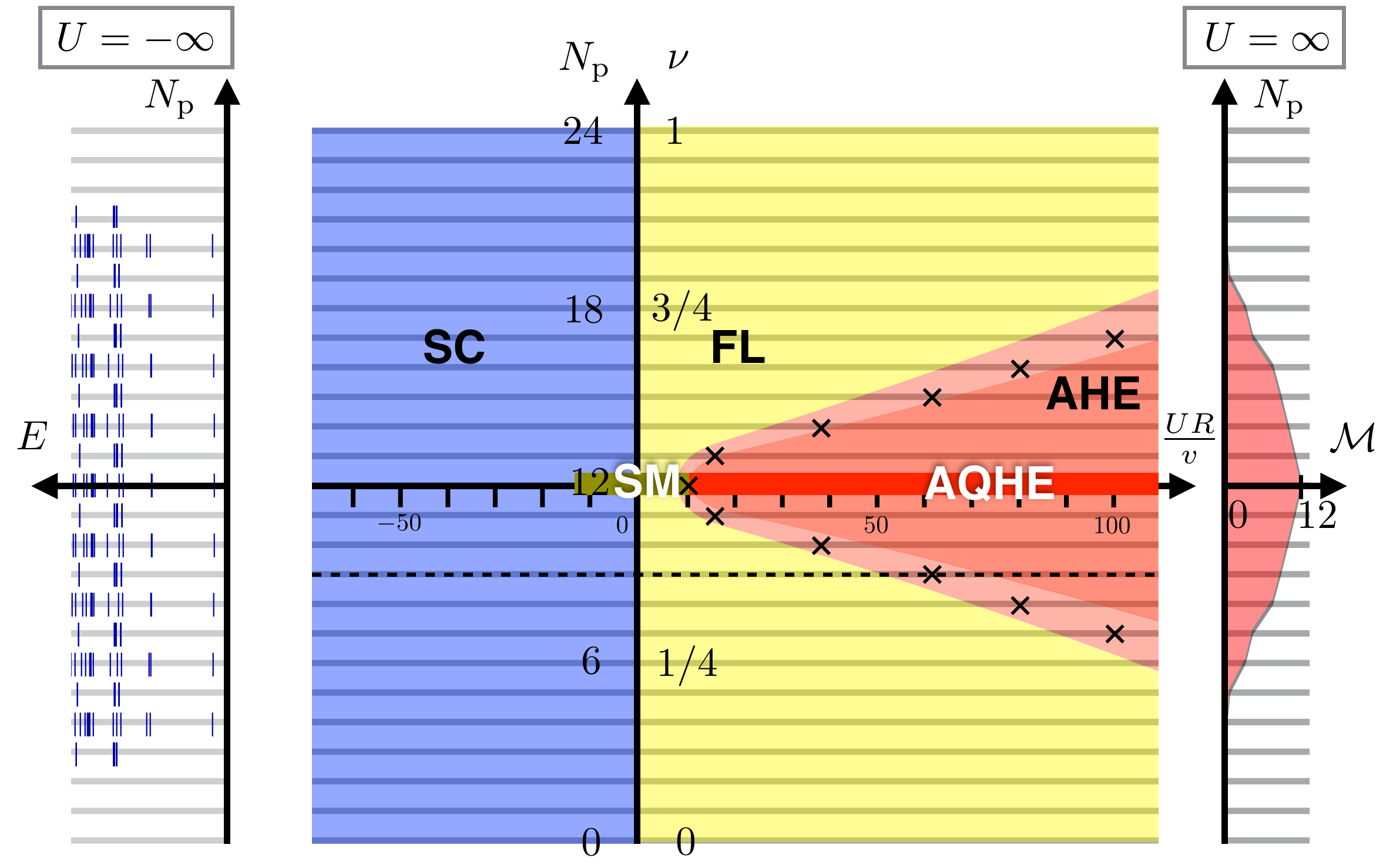}
\caption{(Color online) Phases of the topological insulator surface
  states subject to the contact interaction~\eqref{eq: H int} for a
  Hilbert space restriction $n_0 \le 3$ as a function of interaction
  strength $UR/v$ and filling $\nu=N_{\mathrm{p}}/[2 (n_0+1)(n_0+2)]$.
  Gapped phases are found as an $s$-wave superconductor (SC) and an
  anomalous quantum Hall effect (AQHE) coinciding with
  ferromagnetism. Gapless phases include the semimetal (SM) at half
  filling, a Fermi liquid (FL), and anomalous Hall effect (AHE)
  coinciding with ferromagnetism.  Left panel: Lower end of the energy
  spectrum in the limit $UR/v\to-\infty$ as a function of the particle
  number $N_{\mathrm{p}}$. The superconducting ground state is
  evidenced by the degeneracy of the ground states in all sectors of
  even $N_{\mathrm{p}}$.  Right panel: Magnetization $\mathcal{M}$ of
  the 2-fold (4-fold) quasi-degenerate ground state manifold in the
  limit $UR/v\to\infty$ as a function of the even (odd)
  $N_{\mathrm{p}}$. It evidences spontaneously broken TRS in the
  thermodynamic limit.}
\label{fig:phasedia}
\end{figure}
We have studied the phase diagram of this model as a function of
$UR/v$ and electron filling via exact diagonalization up to $n_0=2$
(24 single particle states) (see Fig.\,\ref{fig:phasedia}).

\begin{figure}[t]
\includegraphics[width=0.98\columnwidth]{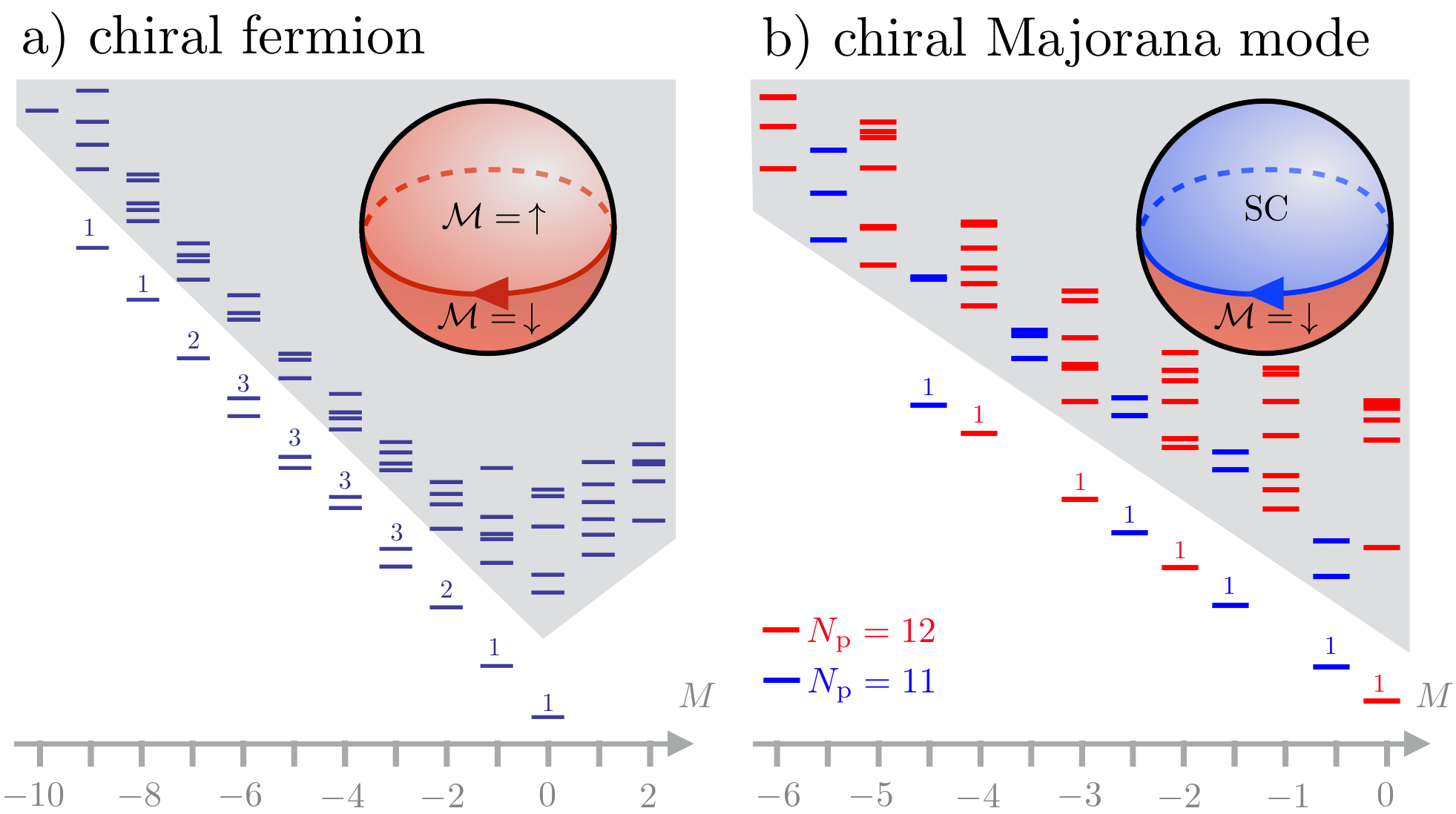}
\caption{(Color online) Numerical evidence for the emergence of (a) 
  a chiral fermionic mode and (b) a chiral Majorana mode at the boundary
  between two ferromagnetic domains and a ferromagnet-superfluid
  domain wall, respectively.  Shown are finite size energy spectra
  for a system restricted to the $n_0\le 2$. Magnetic domains on the
  northern/southern hemisphere are enforced by
  Hamiltonian~\eqref{eq:H0} with a mean-field magnetization
  $m_{\mathrm{z}}(\theta)\sigma_{\mathrm{z}}$ with
  $|m_{\mathrm{z}}|=3.5v/R$, while the superfluid domain is
  interaction-induced with $U=-15v/R$.  In (a), the level counting is
  the one expected for a U(1) mode that consists of six states with
  $m=-5/2,\cdots, 5/2$ assuming that the levels with $m>0$ are
  occupied in the half-filled ground state. In (b),
      the level counting in the fermion parity sectors are as expected
      if one assumes that the chiral Majorana mode at the boundary
      consists of three operators with $m=-5/2,-3/2,-1/2$ which do not
      annihilate the ground state.}
\label{fig:edge modes}
\end{figure}

\begin{figure}[t]
\includegraphics[width=0.98\columnwidth]{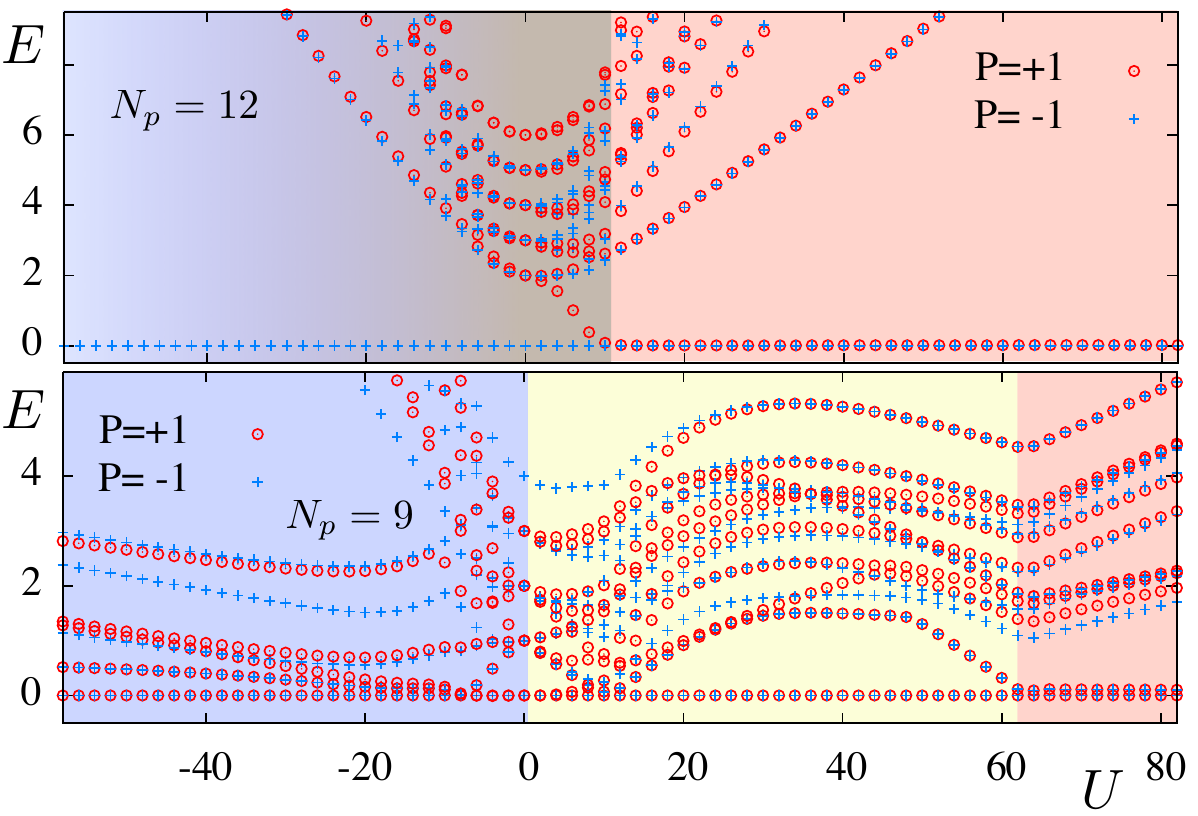}
\caption{(Color online) Exact diagonalization spectra for
  \mbox{$N_p=12$} (top) and $N_p=9$ (bottom) with $n_0=2$ as a
  function of contact interaction $U$.  Together with the numerical
  results for the spin polarization, these spectra lead to the phase
  diagram Fig.\,\ref{fig:phasedia}.  (The color code in both figures
  agrees.)  }
\label{fig:ed}
\end{figure}

{\it Magnetic phases.}---At half filling and for $UR/v>10$, the ground
state is a ferromagnet. In the finite system, we find two
quasi-degenerate ground states 
$|\mathrm{FM}_\pm\rangle$ with $\text{P}=\pm1$ in the $M=0$ sector.
The magnetization operator in $\bs{e}_\text{r}$ direction,
$\Sigma_3\equiv\int_{S_2}\mathrm{d}^2\bs{r}
\left[\rho_\uparrow(\bs{r})-\rho_\downarrow(\bs{r})\right]$,
anticommutes with the parity operator P, since
$\Sigma_3\,\psi^\lambda_{nm}=\,\psi^{-\lambda}_{nm}$.  This implies
that $\langle \mathrm{FM}_+|\Sigma_3|\mathrm{FM}_+\rangle=\langle
\mathrm{FM}_-|\Sigma_3|\mathrm{FM}_-\rangle=0$. The magnetization of
the ferromagnetic ground states with spontaneously broken TRS, which
emerge in the thermodynamic limit, is hence given by
$\mathcal{M}\equiv\langle
\mathrm{FM}_+|\Sigma_3|\mathrm{FM}_-\rangle$.  A ferromagnetically
ordered gapped surface termination of a 3D topological insulator
features a half-integer Hall effect---a phase that would not be
possible in a pure 2D system without intrinsic topological order.
Thus, the ferromagnetic phase also constitutes an anomalous quantum
Hall phase.  Between two domains of opposite magnetization, there
exists a chiral boundary state (see Fig.~\ref{fig:edge modes}a).  Upon
hole- or electron doping the anomalous quantum Hall phase, the system
enters a anomalous Hall phase without a quantized Hall conductance.
At high doping, the ground state is a Fermi liquid which does not
violate any symmetry.  We distinguish these two phases by the
different quasi-degeneracies of the ground state and by computing the
magnetization $\mathcal{M}$ in this quasi-degenerate subspace (see
Fig.~\ref{fig:ed}).~\cite{footnote-finite-size}

{\it Superfluid phase.}---At negative $UR/v$, the system enters a
superfluid phase. For $UR/v\to-\infty$, we find a set of degenerate
states at $M=0$, one in each sector of even particle number
$N_\mathrm{p}$.  Physically, these degeneracies manifest themselves in
the Goldstone mode of the superfluid.  The low energy excitations
above the ground state in each sector of even $N_{\mathrm{p}}$ show
the same structure as the spectrum of two electrons subject to an
infinite repulsive interaction, which consists of three
quasi-degenerate states with $M=-1,0,1$.  This suggests that the
low-energy excitations in the superfluid phase are obtained by
breaking up an individual Cooper pair into two electrons which do not
interact with the condensate.  An $s$-wave superconducting termination
of a 3D topological insulator is a topological superconductor in the
sense that it supports Majorana zero energy states in vortex cores
and a chiral Majorana mode at the boundary with \eg a ferromagnetic
region of the surface (see Fig.~\ref{fig:edge modes}b).  That
we obtain a gapped superconducting state in the limit $UR/v\to-\infty$
is a direct manifestation of the localization properties of the
single-particle states.  If the single particle states were fully
localizable in real space, pairs of electrons could bind into
point-like particles and the ground state would be exponentially
degenerate.

{\it Conclusions.}---We have developed a formalism to study
interaction effects on fermionic 3DTI surface states numerically. From
the analysis of a two-body contact interaction, we found both
ferromagnetic and topologically non-trivial superconducting phases, as
well as chiral fermion and chiral Majorana fermion boundary modes
between different phases. Several branches of future investigation can
be anticipated, such as the application to bosons and studies of more
sophisticated interaction profiles. The formalism establishes an ideal
testing ground for topologically ordered TI surface states scenarios.

\begin{acknowledgments}
  T.N.\ acknowledges several fruitful discussions with F.D.M.\ Haldane
  and thanks him for sharing the results of Ref.~\cite{Haldane14}
  prior to publication. R.T.\ thanks T.\ Dumitrescu and M.\ Metlitski
  for discussions.  This work was supported by DARPA SPAWARSYSCEN
  Pacific N66001-11-1-4110, by the Helmholtz association through
  VI-521, by the DFG through FOR 960 and SPP 1666, and by the European
  Research Council through the grant TOPOLECTRICS
  (ERC-StG-Thomale-336012).
\end{acknowledgments}


\begin{thebibliography}{10}
\bibitem{note}
Note that the sign of the monopole strength is reversed as compared
  to~\cite{Imura12}, since we use a notation where we couple to electrons with
  charge $-e$.
\bibitem{footnote-finite-size}
Phases that become gapless in the thermodynamic limit are hard to detect and
  discriminate using numerially exact diagonalization. Thus, we expect that the
  location of all the phase transition points that we determine as a function
  of $U$ are subject to strong finite-size errors. Since we understand the
  limits at $U=+\infty$, $U=0$ and $U=-\infty$, however, the structure of the
  phase diagram is fixed.
\end{thebibliography}
\end{document}